\documentstyle[preprint,aps,epsf]{revtex}
\begin{document}
\preprint{\parbox{4 cm}{NIKHEF 95-059\\YERPHI-1455(25)-95}}
\thispagestyle{empty}
\title{Longitudinal quark polarization in transversely polarized nucleons}
\author {{\bf A.M. Kotzinian}\footnote{
Now a visitor at CERN, CH-1211, Geneva 23, Switzerland.}\\
{\it Yerevan Physics Institute, Alikhanian Brothers St. 2}\\
{\it AM-375036, Yerevan, Armenia}\\
E-mail: aram@cernvm.cern.ch\\
and\\
{\bf P.J. Mulders}\\
{\it Department of Physics and Astronomy, Free University of Amsterdam}\\
{\it and National Institute for Nuclear Physics and High-Energy Physics}\\
{\it P.O. Box 41882, NL-1009 DB Amsterdam, the Netherlands}\\
E-mail: pietm@nikhef.nl}

\maketitle

\begin{abstract}

Accounting for transverse momenta of the quarks, a longitudinal quark spin
asymmetry exists in a transversely polarized nucleon target. The
relevant leading quark distribution $g_{1T}(x,k_T^2)$ can be measured in
the semi-inclusive deep-inelastic scattering. The average $k_T^2$
weighted distribution function $g^{(1)}_{1T}$ can be obtained directly
from the inclusive measurement of $g_2$.
\end{abstract}

\vspace{1 cm}

Intrinsic transverse momentum ($k_T$) plays an important role in
the quark {\em distribution functions} (DF's) used to describe a polarized
nucleon \cite{rs,tm}.
For the leading (twist-two) part of the deep inelastic scattering cross
section one already needs six DF's to describe the quark state
in a polarized nucleon. They depend on $x$ and $k_T^2$, which
parametrize the quark momentum in a nucleon with momentum $P$,
$k = x\,P + k_T$. We will adopt the notation of Ref.~\cite{tm} for
these "new" six independent DF's: $f^q_1$, $g^q_{1L}$, $g^q_{1T}$, $h^q_{1T}$,
$h_{1L}^{q\perp}$, and $h_{1T}^{q\perp}$ ($q$ denotes the quark flavor).
For a polarized nucleon the spin vector is written as
$S_N$ = $\lambda\,P/M + S_T$, satisfying $\lambda^2 - S_T^2 = 1$.
The probability, ${\cal P}^q_N(x,k_{T}^2)$, the longitudinal spin distribution,
$\lambda^q(x,k_T)$, and the transverse spin distributions,
$s_T^q(x,k_T)$, of the quark in a polarized nucleon are given by
\begin{eqnarray}
&&{\cal P}^q_N(x,k_{T}^2) = f^q_1(x,k_{T}^2), \\
&&{\cal P}^q_N(x,k_{T}^2)\, \lambda^q(x,k_T) =
g^q_{1L}(x,k_T^2)\,\lambda
-g^q_{1T}(x,k_T^2)\,\frac{k_T\cdot S_T}{ M}\,,\\
&&{\cal P}^q_N(x,k_{T}^2) \, s_T^q(x,k_T) \,
= h^q_{1T}(x,k_{T}^2)\,S_T\, + \left[
h_{1L}^{q\perp}(x,k_T^2) \lambda - h_{1T}^{q\perp}(x,k_T^2)
\frac{k_T\cdot S_T}{ M}\right]\,\frac{k_T}{M},
\end{eqnarray}
These DF's have a clear physical interpretation:
for example, $g^q_{1T}$ describes the quark longitudinal polarization
in a transversely-polarized nucleon. Such a polarization can be
nonvanishing only if the quark transverse momentum is nonzero.
This DF cannot be measured in {\it deep-inelastic scattering} (DIS)
at leading order in $1/Q$. It can be measured in polarized
{\it semi-inclusive deep-inelastic scattering} (SIDIS) as first shown
in \cite{ko}, where it appears as an azimuthal asymmetry.
Measurements of the other "new" DF's were proposed in the
doubly-polarized Drell-Yan process \cite{rs,tm},
and in the polarized SIDIS \cite{ko,tm2} using the so called
Collins effect \cite{col}. The quark fragmentation is
described by two {\it fragmentation functions} (FF's): spin--independent
and transverse--spin--dependent ones.

The "ordinary", $f^q_1(x)$, $g^q_1(x)$ and $h^q_1(x)$, and the "new"
leading-twist DF's are related by $k_T$--integration
\begin{eqnarray}
&&f^q_1(x)=\int d^{\,2}k_T \,f^q_1(x,k_{T}^2),\\
&&g^q_1(x)=\int d^{\,2}k_T \,g^q_{1L}(x,k_{T}^2),\\
&&h^q_1(x)=\int d^{\,2}k_T \left[h^q_{1T}(x,k_{T}^2)-
\frac{k_T^2}{2 M^2}h_{1T}^{q\perp}(x,k_{T}^2)\right].
\end{eqnarray}
The DF $g^q_{1T}(x,k_T^2)$ does not contribute to $g^q_1(x)$,
but it does contribute to the DF $g^q_T(x)$ = $g^q_1(x)+g^q_2(x)$,
which contributes at ${\cal O}(1/Q)$ in the inclusive polarized
leptoproduction cross section \cite{tm1}.
A detailed discussion of the DF $g^q_2$ is given in
the recent review by Anselmino, Efremov and Leader \cite{ael}.

In this letter we will be mainly concerned with the longitudinal quark
spin distribution $\lambda^q(x,k_T)$ and the two DF's
$g^q_{1L}(x,k_T^2)$ and $g^q_{1T}(x,k_T^2)$ describing it.
Following Ref.~\cite{ko}, we first consider the polarized SIDIS
in the simple quark-parton model. We will use the standard
notation for DIS variables: $l$ and $l'$ are the momenta of the initial
and the final state lepton; $q=l-l'$ is the exchanged virtual photon
momentum; $P$ ($M$) is the target nucleon momentum (mass), $S$ its spin;
$P_h$ is the final hadron momentum; $Q^2=-q^2$; $s=Q^2/xy$;
$x=Q^2/2P\cdot q$; $y=P\cdot q/P\cdot l$; $z=P\cdot P_h/P\cdot q$.
The reference frame is defined with the $z$-axis along the virtual
photon momentum direction (antiparallel) and
$x$-axis in the lepton scattering plane, with positive direction chosen along
lepton transverse momentum. Azimuthal angles of the produced hadron,
$\phi_h$, and of the nucleon spin, $\phi_S$, are counted around $z$-axis
 (for more details see Refs \cite{ko} or \cite{mt}).
In this letter as independent azimuthal angles we will choose
$\phi_h^S$ $\equiv$ $\phi_h - \phi_S$ and
$\phi_l^S$ $\equiv$ $\phi_l - \phi_S$ and we will give
cross-sections integrated over $\phi_l^S$ at fixed value of $\phi_h^S$.

In leading order in $1/Q$ the SIDIS cross section for polarized
leptons and hadrons has the form
\begin{equation}
\frac{d\sigma (\ell N\rightarrow \ell^\prime hX)}
{dxdydz\,d^{\,2}P_{h\perp}}=
\frac{2\pi\alpha^2}{Q^2\,y}\,[1+(1-y)^2]\, \left\lgroup
{\cal H}_{f_1}^0+D(y)
\,\biggl[ \lambda\, {\cal H}_{g_{1L}}^0+
\vert S_T\vert\, \cos\phi_h^S\,{\cal H}_{g_{1T}}^0\biggr]\right\rgroup,
\label{sig}
\end{equation}
where
\begin{equation}
D(y) = \frac{y(2-y)}{1 + (1-y)^2}
\end{equation}
is the depolarization of the virtual photon with respect to the parent lepton.
We do not consider here the
cross section for unpolarized leptons and polarized hadrons which involves
the structure functions ${\cal H}_{h_{1T}}^S$, ${\cal H}_{h_{1L}^\perp}^S$,
and ${\cal H}_{h_{1T}^\perp}^S$ \cite{ko,tm2}.
This single-polarized part of the SIDIS cross-section drops out
after integration over $\phi^S_l$ in leading order in $1/Q$.

The structure functions ${\cal H}_{f}^0$ entering in Eqs (\ref{sig})
are given by quark-charge-square
weighted sums of definite $k_T$-convolutions of the DF's and
the well-known spin-independent FF $D_q^h(z,(P_{h\perp}-zk_T)^2)$.
Taking into account the transverse momentum the latter depends on $z$ and
the transverse momentum squared of the produced hadron relative to the
parent quark. Neglecting radiative corrections, the functions
are independent of $Q^2$, however.
The explicit form of the structure functions
can be found in refs \cite{ko} or \cite{mt}:
\begin{eqnarray}
{\cal H}_{f_1}^0 &=& \sum_q e_q^2 \,\int d^{\,2}k_T
\,f^q_1(x,k_T^2) D_q^h(z,(P_{h\perp}-zk_T)^2),\\
{\cal H}_{g_{1T}}^0 &=& \sum_q e_q^2 \,\int d^{\,2}k_T
\frac{\bbox{k}_T\cdot \bbox{P}_{h\perp}}
{M\,\vert \bbox{P}_{h\perp}\vert}
\,g^q_{1T}(x,k_T^2) D_q^h(z,(P_{h\perp}-zk_T)^2),\\
{\cal H}_{g_{1L}}^0 &=& \sum_q e_q^2 \,\int d^{\,2}k_T
\,g^q_{1}(x,k_T^2) D_q^h(z,(P_{h\perp}-zk_T)^2).
\label{hpm}
\end{eqnarray}
Note, that these structure functions include only the
rather well studied unpolarized FF's, $D_q^h(z)$.

The target-longitudinal-polarization asymmetry is defined as
\begin{equation}
{\cal A}_{L}(x,y,z,P_{h\perp})
=\frac{d\sigma^{\rightarrow}-d\sigma^{\leftarrow}}
{d\sigma^{\rightarrow}+d\sigma^{\leftarrow}},
\label{aldef}
\end{equation}
where $\rightarrow$ ($\leftarrow$) means longitudinal polarization,
$\lambda$ = 1 (-1) and $S_T$ = 0.
Analogously, the target-transverse-spin asymmetry is defined as
\begin{equation}
{\cal A}_{T}(x,y,z,P_{h\perp},\phi^S_h)
=\frac{d\sigma^{\uparrow}-d\sigma^{\downarrow}}
{d\sigma^{\uparrow}+d\sigma^{\downarrow}}.
\label{atdef}
\end{equation}
with $\uparrow$ ( $\downarrow$ ) denoting the transverse
polarization of the target nucleon with $\lambda$ = 0 and
$\vert \bbox{S}_T \vert$ = 1.

The phase space element in the transverse direction is
$d^{\,2} P_{h\perp}$= $\vert \bbox{P}_{h\perp}\vert \,d\,\vert
\bbox{P}_{h\perp}\vert\,d\phi^S_h$.
Integrating ${\cal A}_L$
over $\phi^S_h$ we are left with the contribution proportional
to ${\cal H}_{g_{1L}}^0$,
\begin{equation}
\int \frac{d\phi^S_h}{2\pi}\,{\cal A}_L
= D(y) \,\frac{{\cal H}_{g_{1L}}^0}{{\cal H}_{f_1}^0}.
\end{equation}
One can also define the asymmetry
\begin{equation}
\langle A_{L} \rangle\,(x,y,z) \equiv \frac{\int d^{\,2}P_{h\perp}
(d\sigma^{\rightarrow}-d\sigma^{\leftarrow})}
{\int d^{\,2}P_{h\perp}(d\sigma^{\rightarrow}+d\sigma^{\leftarrow})}
=D(y)\,\frac{\sum_q e_q^2\, g^q_1(x)\,D_q^h(z)}
{\sum_q e_q^2\,f^q_1(x) \,D_q^h(z)}.
\label{alm}
\end{equation}
This asymmetry was measured by the SMC collaboration \cite{smc} and
provides the flavour analysis of the quark longitudinal-spin
DF's in longitudinally polarized nucleon \cite{close-milner,nnn}.
The future measurements are planned by the HERMES \cite{herm} and the
HMC \cite{hmc} collaborations.

The target-transverse-spin asymmetry is given by
\begin{equation}
{\cal A}_{T}(x,y,z,P_{h\perp},\phi^S_h)=
D(y)\,\cos\phi_h^S\,\frac{{\cal H}_{g_{1T}}^0}{{\cal H}_{f_1}^0}
\label{ath}
\end{equation}
and can in principle be disentangled measuring the asymmetry at different
values of $\phi^S_h$ and performing a Fourier analysis. For
example, let us integrate Eq.~\ref{ath} weighted by $\cos\phi_h^S$
over $\phi^S_h$. We obtain
\begin{equation}
\int_0^{2\pi} \frac{d\phi^S_h}{2\pi}\,\cos\phi_h^S
\,{\cal A}_{T}(x,y,z,P_{h\perp},\phi^S_h) =\frac{1}{2}
D(y) \frac{{\cal H}_{g_{1T}}^0}{{\cal H}_{f_1}^0}.
\label{at1}
\end{equation}
It is useful to define the transverse-spin asymmetry weighted with
$\bbox{S}_T\cdot \bbox{P}_{h\perp}/M$ =
$(\vert \bbox{P}_{h\perp}\vert/M)\cos\phi^S_h$,
\begin{eqnarray}
\langle \frac{\vert\bbox{P}_{h\perp}\vert}{M}\cos\phi_h^S\,
A_{T} \rangle\,(x,y,z) & = & \frac{
\int d^{\,2}P_{h\perp} \frac{\vert
\bbox{P}_{h\perp}\vert}{M}\,\cos\phi_h^S\,
(d\sigma^{\uparrow}-d\sigma^{\downarrow})}
{\int d^{\,2}P_{h\perp}(d\sigma^{\uparrow}+d\sigma^{\downarrow})}
\nonumber \\ & = &
zD(y)\,\frac{\sum_q
e_q^2\, g^{q\,(1)}_{1T}(x)\,D_q^h(z)} {\sum_q e_q^2\,f^q_1(x) \,D_q^h(z)},
\label{atm}
\end{eqnarray}
where
\begin{equation}
g_{1T}^{q\,(1)}(x) = \int d^{\,2}k_T\,\frac{\bbox{k}_T^2}{2M^2}
\,g^q_{1T}(x,k_T^2).
\end{equation}
Note, that relations \ref{alm} and \ref{atm} are
valid for any $k_T$--dependence of DF's and FF's.
In principle, it is possible to separate contributions from
different quark flavours by measuring the asymmetry \ref{atm}
for different produced hadrons in the same way as proposed in
\cite{close-milner,nnn}.

\begin{figure}
\begin{center}
\leavevmode
\epsfxsize=9 cm
\epsfbox{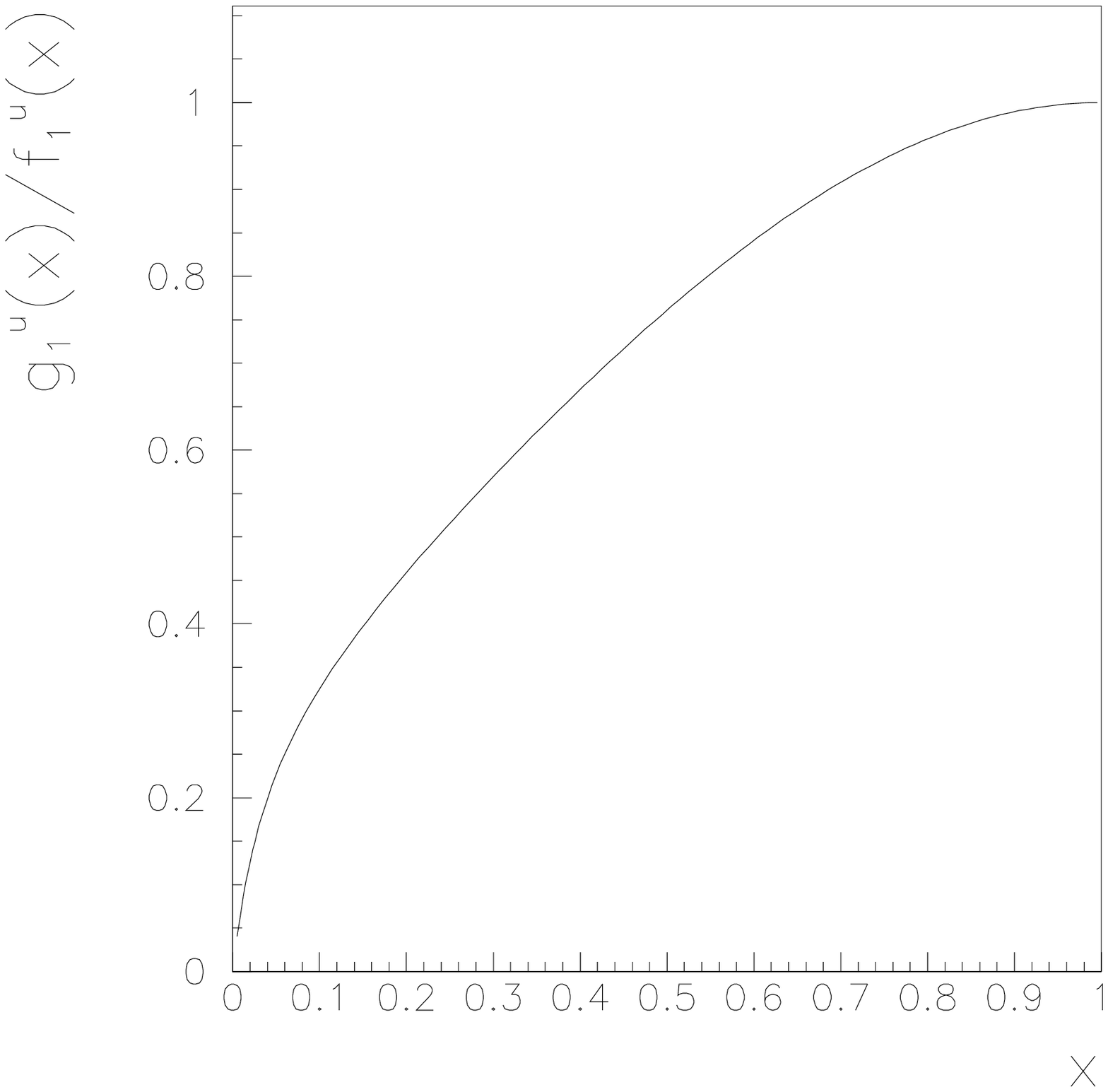}
\end{center}
\caption{\label{fig1}
The longitudinal spin asymmetry (Eq.~\protect\ref{longas})
as function of $x$ with BBS-parametrization.}
\end{figure}

Next we turn to a quantitative estimates of the asymmetries,
starting with the longitudinal asymmetry. For this we consider the
production of $\pi^+$-mesons on the proton. The dominant contribution will
come from scattering on the $u$-quark. In order to estimate $\langle
A_L\rangle$ we use the parametrization of Brodsky, Burkardt and
Schmidt (BBS) \cite{bbs} for $g^u_1$ and $f^u_1$. The result,
\begin{equation}
\frac{1}{D(y)}\,\langle A_L\rangle\,(x,y,z)
\approx \frac{g_1^u(x)}{f_1^u(x)},
\label{longas}
\end{equation}
is shown in Fig.~\ref{fig1}.

For an estimate of the transverse asymmetry we need the DF's
$g_{1T}^q(x,k_T^2)$. In contrast to the $k_T$-integrated DF's $f_1^q(x)$ and
$g_1^q(x)$ there is no measurements of the function $g_{1T}^q(x,k_T^2)$.
As is shown in Ref.~\cite{tm1,mt} the
$(\bbox{k}_T^2/2M^2)$--weighted $k_T$-integrated function
$g_{1T}^{q\,(1)}(x)$, which appears in Eq.~\ref{atm} is directly
related to the DF $g_2^q(x)$,
\begin{equation}
g^q_2(x) = \frac{d}{dx}\,g^{q\,(1)}_{1T}.
\label{gt2}
\end{equation}
This relation just follows from constraints imposed by Lorentz
invariance on the antiquark-target forward scattering amplitude and
the use of QCD equations of motion for quark fields.
We will use this relation for our quantitative estimates. We note that
the effects of higher order QCD corrections for the transverse momentum
dependent functions, however, require further investigation
\cite{Dokshitzer}. For the function $g_2$ the QCD corrections have
been extensively studied \cite{collect}.

\begin{figure}[tb]
\begin{center}
\leavevmode
\epsfxsize=9 cm
\epsfbox{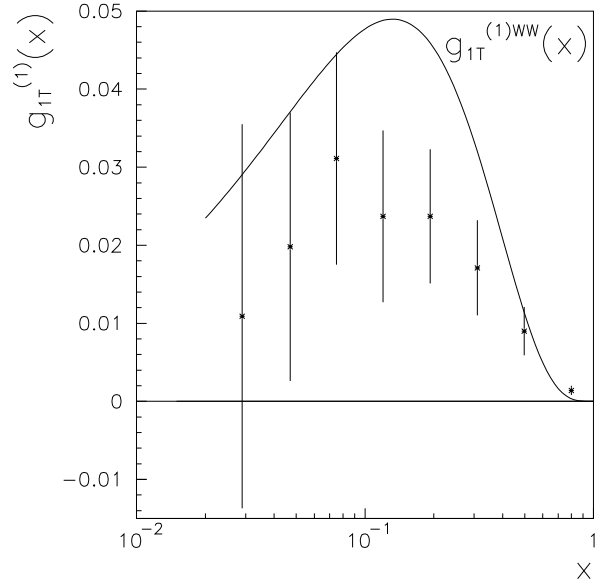}
\end{center}
\caption{\label{fig2}
The function $g_{1T}^{(1)}(x)$ as obtained from E143 data
using Eq.~\protect\ref{g11tg2} or from the BBS-parametrizations for $g_1$ using
Eq.~\protect\ref{g11tww}.}
\end{figure}

In our first estimate for the transverse asymmetry (Eq.~\ref{atm})
we use recent data on $g_2$ and the relation in Eq.~\ref{gt2}.
Such data are available from the SMC collaboration \cite{smc1}
and the E143 collaboration at SLAC \cite{E143}.
Particularly the latter data allow a rough estimate of the function,
\begin{equation}
g_{1T}^{(1)}(x) = \frac{1}{2}\,\sum_q e_q^2 g_{1T}^{q\,(1)}(x) =
\int d^{\,2}k_T \frac{\bbox{k}_T^2}{2M^2}\,g_{1T}(x,k_T^2) =
- \int_x^1 dy\,g_2(y).
\label{g11tg2}
\end{equation}
The result obtained by averaging the two sets of data at different
angles and adding statistical and systematic errors quadratically,
is shown in Fig.~\ref{fig2}.

Our second estimate for this distribution function comes from the
representation for $g_2^q(x)$ in terms of other $k_T$-integrated
functions. For $g_T^q$ = $g_1^q+g_2^q$ one has
\begin{equation}
g_T^q(x) = \int_x^1 dy\frac{g_1^q(y)}{y}+
\frac{m_q}{M}\bigg[\frac{h_1^q(x)}{x}-
\int_x^1dy\frac{h_1^q(y)}{y^2}\bigg]+
\tilde{g}^q_T(x)-\int_x^1dy\frac{\tilde{g}^q_T(y)}{y},
\label{gt1}
\end{equation}
where $m_q$ is the quark mass, and $\tilde{g}^q_T$ is
the so called interaction-dependent part of the DF $g^q_T(x)$.
The term $(m_q/Mx)h_1^q(x)$ in the {\it rhs} of Eq.~\ref{gt1} was
found many years ago by Feynman \cite{feynman} and represents the
contribution of the transverse spin distribution to $g_T(x)$. The most
well-known contribution in Eq.~\ref{gt1} is the first term found by Wandzura
and Wilczek \cite{ww}.
Using Eq.~\ref{gt2} an estimate of $g_{1T}^{(1)}(x)$ is obtained from
Eq.~\ref{gt1}, keeping only the first term (Wandzura-Wilczek) as this does
not contradict the data. In that case
$g_T(x) = g_T^{WW}(x)$ where
\begin{equation}
g_T^{WW}(x) = g_1(x) + g_2^{WW}(x) = \int_x^1 dy \,\frac{g_1(y)}{y},
\end{equation}
leading to
\begin{equation}
g_{1T}^{(1) WW}(x) = - \int_x^1 dy \,g_2^{WW}(y)
= x\int_x^1 dy\,\frac{g_1(y)}{y} = x\,g_T^{WW}(x).
\label{g11tww}
\end{equation}
We use the parametrization of DF's from Ref.~\cite{bbs}.
The result is shown as the curve in Fig.~\ref{fig2}.
Using other parametrizations for $g_1$ does not substantially
change this result.

\begin{figure}[tb]
\begin{center}
\leavevmode
\epsfxsize=9 cm
\epsfbox{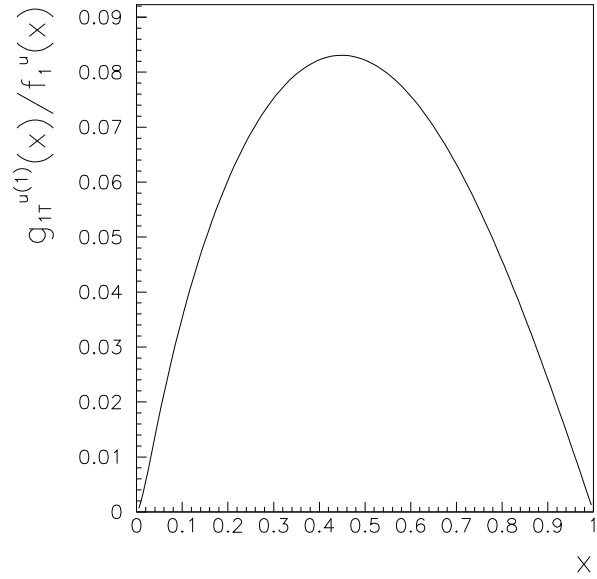}
\end{center}
\caption{\label{fig3}
The transverse spin asymmetry (Eq.~\protect\ref{tras})
as function of $x$ estimated from the BBS-parametrization for $g_1$
using Eq.~\protect\ref{g11tww}.}
\end{figure}

Assuming the $u$-quark dominance for the $\pi^+$ production on
the proton, the estimate for the transverse spin asymmetry,
\begin{equation}
\frac{1}{z\,D(y)}\,\langle
\frac{\vert \bbox{P}_{h\perp}\vert}{M}\,\cos\phi_h^S
\, A_T \rangle\,(x,y,z) \approx \frac{g^{u\,(1)}_{1T}(x)}{f^u_1(x)},
\label{tras}
\end{equation}
can be obtained (see Fig.~\ref{fig3}).

In this letter we have considered the azimuthal asymmetry in 1-particle
inclusive polarized leptoproduction.
The longitudinal spin asymmetry averaged over the
transverse momenta of the produced hadrons gives independent ways to study
the polarized quark distributions as has been pointed out before
\cite{close-milner,nnn}.
As we have shown the transverse spin asymmetry provides
information on the quark-longitudinal spin distribution in a
transversely polarized target, the DF $g^q_{1T}(x,k_T^2)$.
This information appears in a $\cos (\phi_h - \phi_S)$
asymmetry for the produced particles. A Fourier analysis of this asymmetry
weighted with the modulus of the transverse momentum of produced particles,
gives the $k_T$--integrated and $\bbox{k}_T^2/2M^2$--weighted function
$g^{q\,(1)}_{1T}(x)$ which is at tree-level directly related to $g^q_2(x)$.
This provides an alternate way of obtaining the latter DF, although a
careful analysis of the QCD corrections is needed.

A.K. is grateful to A.V. Efremov and Yu.I. Dokshitzer for useful
discussions.
Part of the work of P.M. was supported by the
foundation for Fundamental Research on
Matter (FOM) and the Dutch National Organization for Scientific Research
(NWO).

\newpage

\end{document}